\documentclass[graybox, envcountchap]{svmult}

\usepackage{mathptmx}        % selects Times Roman as basic font
\usepackage{helvet}          % selects Helvetica as sans-serif font
\usepackage{courier}         % selects Courier as typewriter font
\usepackage{makeidx}         % allows index generation
\usepackage{graphicx}        % standard LaTeX graphics tool
\usepackage{multicol}        % used for the two-column index
\usepackage[bottom]{footmisc}% places footnotes at page bottom

\makeindex             % used for the subject index
\usepackage{amsmath}
\usepackage{amssymb}
\usepackage{lscape}
\usepackage{multirow}

\def\gapp{\ifmmode\stackrel{>}{_{\sim}}\else$\stackrel{<}{_{\sim}}$\fi}
\def\gsim{\lower.5ex\hbox{\gtsima}}
\def\gtsima{$\; \buildrel > \over \sim \;$}
\def\lapp{\ifmmode\stackrel{<}{_{\sim}}\else$\stackrel{<}{_{\sim}}$\fi}
\def\lsim{\lower.5ex\hbox{\ltsima}}
\def\ltsima{$\; \buildrel < \over \sim \;$}

\newcommand\apgt{\ {\raise-.5ex\hbox{$\buildrel>\over\sim$}}\ }
\newcommand\aplt{\ {\raise-.5ex\hbox{$\buildrel<\over\sim$}}\ }
\newcommand{\msun}{{\ensuremath {\rm M}_\odot}}

\begin{document}
\pagestyle{empty}
\frontmatter%%%%%%%%%%%%%%%%%%%%%%%%%%%%%%%%%%%%%%%%%%%%%%%%%%%%%%

\include{dedic}
\include{foreword}
\include{preface}

\mainmatter%%%%%%%%%%%%%%%%%%%%%%%%%%%%%%%%%%%%%%%%%%%%%%%%%%%%%%%

\setcounter{chapter}{11}

\title{Models of Individual Blue Stragglers}

\author{Alison Sills}

\institute{Alison Sills \at Department of Physics and Astronomy, McMaster University, 1280 Main Street West, Hamilton, ON, L8S 4M1, CANADA, \email{asills@mcmaster.ca}}

\maketitle

\label{Chapter:Sills}
%Use \abstract*{} if abstract is NOT to appear in book, otherwise remove the *.

\abstract*{This chapter describes the current state of models of individual blue stragglers. Stellar collisions, binary mergers (or coalescence), and partial or ongoing mass transfer have all been studied in some detail. The products of stellar collisions retain memory of their parent stars and are not fully mixed. Very high initial rotation rates must be reduced by an unknown process to allow the stars to collapse to the main sequence. The more massive collision products have shorter lifetimes than normal stars of the same mass, while products between low mass stars are long-lived and look very much like normal stars of their mass. Mass transfer can result in a merger, or can produce another binary system with a blue straggler and the remnant of the original primary. The products of binary mass transfer cover a larger portion of the colour-magnitude diagram than collision products for two reasons: there are more possible configurations which produce blue stragglers, and there are differing contributions to the blended light of the system. The effects of rotation may be substantial in both collision and merger products, and could result in significant mixing unless angular momentum is lost shortly after the formation event. Surface abundances may provide ways to distinguish between the formation mechanisms, but care must be taking to model the various mixing mechanisms properly before drawing strong conclusions. Avenues for future work are outlined.}

\section{Introduction}
\label{sec:Introduction}

As we have seen in previous chapters, blue stragglers are thought to form in both of two broad classes of formation mechanisms: through collisions\index{collision} or through binary mass transfer\index{mass transfer}. Both processes transform two stars into a blue straggler on reasonably short timescales compared to the main sequence lifetimes of low mass stars: a head-on collision can take a few days, while stable mass transfer from main sequence stars can take much longer, up to the nuclear timescale of the stars. However, it is not enough to follow the hydrodynamic phases of either process. We must determine the structure of the proto-blue straggler at the moment when the remnant is in hydrostatic equilibrium, and then follow its evolution using a stellar evolution code\index{stellar evolution code} for the next few billion years in order to make direct comparisons to the observed properties of blue stragglers. Those stellar evolution models are the subject of this chapter.

In the context of stellar models, it makes sense to try to define these two formation mechanisms a little more clearly. For our purposes, a collision is a short-lived and strong interaction between two stars. The situation where two previously unrelated stars move through a cluster and find themselves close enough that their radii overlap is obviously a collision, and those could be either head-on or off-axis. A similar situation during a resonant encounter\index{resonant encounter} involving binaries or triples can also be modelled using the same techniques as a collision. In this case the two stars might have originally been a binary, but the event that made the blue straggler was not a slow spiral-in of the two stars, but a more dramatic event caused by the overall interaction of the systems. In the literature, these kinds of collisions are often called ``binary-mediated". The key distinction that I would like to make is that the event which triggered the hydrodynamic phase of a collision is very short-lived, and determined by the environment in which the star finds itself. A binary mass transfer blue straggler, on the other hand, has typically gained its mass due to the orbital and nuclear evolution of the stars in a binary (or triple) system. The orbit of the system may have been modified by interactions in a cluster earlier in the binary's life, but the event which caused the system to begin transferring mass is intrinsic to the binary. 

These distinctions between the two formation mechanisms drive the ways in which the blue stragglers can be modelled. A collision is a short, single event which takes two (or three) stars and transforms them into something else -- but when that event is over, the star(s) return to thermal equilibrium\index{thermal equilibrium} and can be modelled using a normal stellar evolution code\index{stellar evolution code}. The initial configuration of the gas may not be the same as a standard single star, but there will not be any more externally-imposed changes to the system. In binary mass transfer, however, the changes to the structure of the star could be happening at the same time as more mass is being added to the blue straggler. Mass transfer can happen on dynamic timescales\index{dynamical timescale}, during which a large amount of mass is dumped onto the accretor. If the outcome of this dynamical mass transfer event is a fully detached and non-interacting binary, or the complete destruction of the donor, then the subsequent evolution of the blue straggler can be treated in a similar way that we treat the evolution of collision products. These outcomes of mass transfer are extremely unlikely, however. Typically, either a dynamical mass transfer event results in a common envelope\index{common envelope} or contact binary\index{contact binary} situation; or the mass transfer is stable and takes place on timescales that are comparable to the nuclear\index{nuclear timescale} or thermal timescale\index{thermal timescale} of the star. In these cases, a standard single-star stellar evolution code is not sufficient to follow the evolution of the blue straggler and other methods are required. 

In the following sections I will introduce the various methods that have been used to model both collision and binary mass transfer blue stragglers. I will outline the successes of each approach in describing particular observations, and will also discuss areas where the models are inconclusive or incomplete. In the final sections, I will describe observations of blue stragglers which could be compared to detailed stellar models, but for which the models are just starting to be available. 

\section{Collisional Models}
\label{sec:CollisionalModels}

The first hydrodynamical simulations\index{hydrodynamical simulation} of collisions\index{collision} between main sequence stars were done in the late 1980s \cite{BenzHills87}. These low resolution simulations (1024 particles) were interpreted by the authors to show that the collision products were fully mixed. Based on these results, the first stellar evolution calculations of stellar collisions were done by \cite{BailynPinsonneault95}. They took normal stars with masses appropriate for blue stragglers in globular clusters (0.8 to 1.8 $\msun$) and evolved them with helium abundances up to $Y=0.315$. In the same paper, the authors also approximated binary mass transfer blue stragglers as chemically inhomogeneous stars, as if they were starting their lives part-way along the normal main sequence track. Their conclusion that the outer blue stragglers in the cluster M3\index{M3} are inconsistent with the collisional hypothesis, and are well-fit by the binary merger tracks needs to be re-evaluated in the light of a subsequent change in our understanding of the evolution of stellar collision products.

A decade after the Benz \& Hills \cite{BenzHills87} results, higher resolution Smoothed Particle Hydrodynamics\index{Smoothed Particle Hydrodynamics} (SPH) simulations of main sequence stellar collisions were performed \cite{Lombardi95,Lombardi96} which showed that the earlier interpretation of full mixing was incorrect and that those authors had been misled by their low resolution. In fact, collision products retain a strong memory of their parent stars, with the cores of the more evolved stars ending up in the core of the collision product, and the less evolved star ends up on the outside of the collision product. Based on these simulations, Procter Sills, Bailyn, \& Demarque \cite{Procter95} reworked the simulations of \cite{BailynPinsonneault95} by mapping a composition profile of an evolved low-mass star onto a higher mass star, and concluded that unmixed blue stragglers would have lifetimes that are very short, too short to explain the brightest and bluest blue stragglers in some globular clusters, particularly NGC 6397\index{NGC 6397}. 

Sills \& Lombardi \cite{SillsLombardi97} realised that the structure of the collision product could be simply approximated by sorting the fluid of the two parent stars by their entropy\index{entropy}. A stable star has entropy increasing outwards, and so we can simply compare the entropy of each shell in the two stars and place them in order from lowest to highest. Since material that has a higher mean molecular weight has a lower entropy, any material that is helium-rich will fall to the centre of the collision product. This ``sort by entropy" prescription is the basis for the codes {\tt Make Me A Star}\index{Make Me A Star code} ({\tt MMAS}) \cite{MMAS} and {\tt Make Me A Massive Star}\index{Make Me A Massive Star code} ({\tt MMAMS}) \cite{MMAMS}, which provide detailed stellar structure profiles of collision products for collisions between low mass and high mass stars respectively. 

At the end of the hydrodynamic simulations\index{hydrodynamic simulation}, we are left with a ball of gas that is in hydrostatic equilibrium\index{hydrostatic equilibrium} but not necessarily in thermal equilibrium\index{thermal equilibrium}. This configuration can be modelled by most stellar evolution codes\index{stellar evolution code}, as long as the object is not too far out of thermal equilibrium. While ``too far" is not easily quantified, the endpoints of stellar collision calculations can be used as the starting models for stellar evolution codes, with only slight modifications to treat the outer boundary conditions. Most SPH simulations (and also {\tt MMAS/MMAMS}) do not have the resolution to accurately treat the very tenuous atmosphere of a star, so some extrapolation needs to be done to make a complete stellar model. However, the internal pressure, temperature, density, rotation rate, and composition of the collision product can be imported directly. This is an improvement over the early, more ad hoc models, and provides us with more certainty that we are following the evolution of a real collision product for many gigayears after the collision.

\begin{figure}
%\sidecaption
\includegraphics[width=119mm]{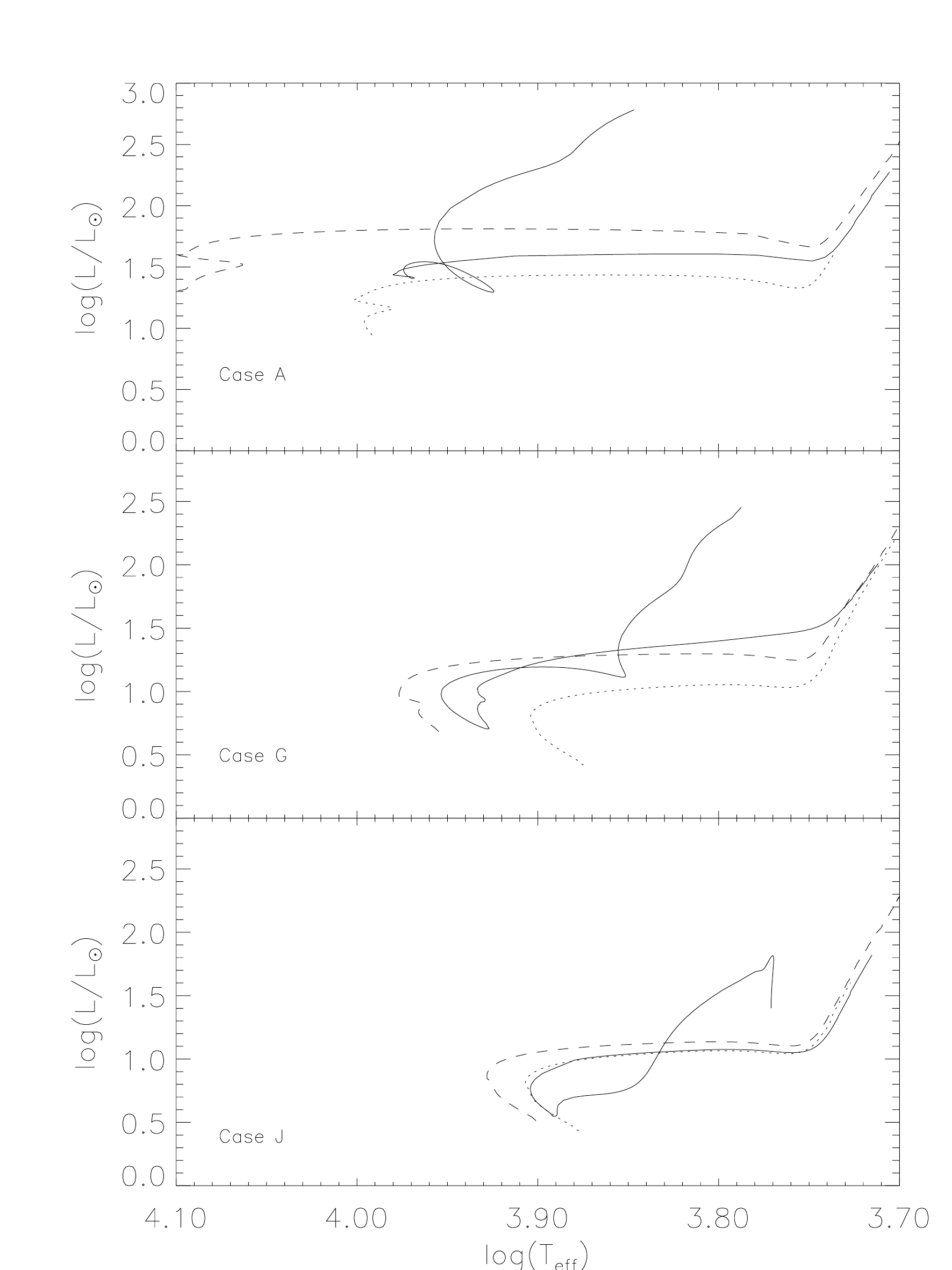}
\caption{Evolutionary tracks for a collision product (solid line), a star with the same mass and the composition of a normal globular cluster star (dotted line), and a version of the collision product which was fully mixed before evolution was started (dashed line). The normal and fully mixed evolutionary tracks are shown only from the zero age main sequence onwards. Case A is a collision between two turnoff-mass (0.8 $\msun$) stars; case G is a collision between a 0.8 $\msun$ star and a 0.4 $\msun$ star, and case J is a collision between two 0.6 $\msun$ stars. (Figure 6 from \cite{Sills97}, reproduced by permission of the AAS) }
\label{fig:Collisions}
\end{figure}

After the high resolution SPH simulations were published, a number of groups combined hydrodynamics with stellar evolution to model blue stragglers \cite{Sandquist97,Sills97,Ouellette98}. All groups were interested in the properties of the collision product during the thermal-timescale contraction back to the main sequence. This phase is somewhat analogous to the pre-main sequence phase, except that the blue straggler is not chemically homogeneous and therefore does not follow the Hayashi track\index{Hayashi track}. The other major difference, found by all three groups, is that the collision product is stable against convection throughout this contraction phase. Therefore, there is no mixing of material into the core of the star even after the collision. As a result, collisions that involved a star at or near the turnoff in a cluster has a core which is depleted in hydrogen and therefore has a short main sequence lifetime. Collisions between stars further down the main sequence have less helium in their cores and therefore have longer lifetimes. In general, however, the evolutionary track of a blue straggler is very close to that of a normal star of the same mass. If the collision product has a helium-enriched core, it acts as if it is starting its life somewhere between the zero-age and the terminal-age main sequence. Evolutionary tracks of head-on collision products are shown in Fig.~\ref{fig:Collisions}, with a comparison of tracks for normal stars of the same mass as well. These techniques were extended to stars of solar metallicity \cite{Glebbeek08a,Glebbeek08b}, and those studies also investigated the effects of modifying the ages of the parent stars. The globular cluster studies assumed that the stars that collided were the same age as the cluster, but there is nothing special about the current time in any given cluster and collision products can have reasonably long lifetimes. Therefore, it is important to look at the difference between collision products that happened at various times in the past. Based on these detailed evolution calculations, Glebbeek \& Pols \cite{Glebbeek08b} presented simple formulae for the lifetimes, luminosities, temperatures, and radii of collision products compared to normal stars of the same mass. 

Most of the models discussed above assumed that the collisions between stars were exactly head-on\index{head-on collision}. This is extremely unlikely to actually occur, so models of off-axis collisions\index{off-axis collision} were also calculated by \cite{Lombardi96} and evolved by \cite{Sills01}, and also investigated by \cite{Ouellette98}. In general, the density, pressure, temperature and chemical composition profiles of a collision product do not depend much on the impact parameter of the collision (where an impact parameter of 0 is a head-on collision, and a grazing collision would have an impact parameter of 1.0 in units of the sum of the initial stellar radii $R_1 + R_2$). These quantities are dictated more by the structure of the parent stars. However, the angular momentum of the collision product increases significantly as the impact parameter increases. In all SPH simulations, the parent stars are initially not rotating. However, there is orbital angular momentum\index{angular momentum} in any collision that is not directly head-on. Since the two stars become one object, the bulk of this angular momentum must end up in the collision product. Typically 1-6\% of the total mass of the system is unbound during the collision, so some angular momentum can be lost this way. However, the specific angular momentum at the surface of the parent stars is very low, and off-axis collisions lose less mass than the head-on collisions. Therefore, the collision products are rotating quite rapidly immediately after the collision. The total angular momentum of the collision products can be as much as ten times higher than normal pre-main sequence stars of comparable mass. 

\begin{figure}
\sidecaption
\includegraphics[width=119mm]{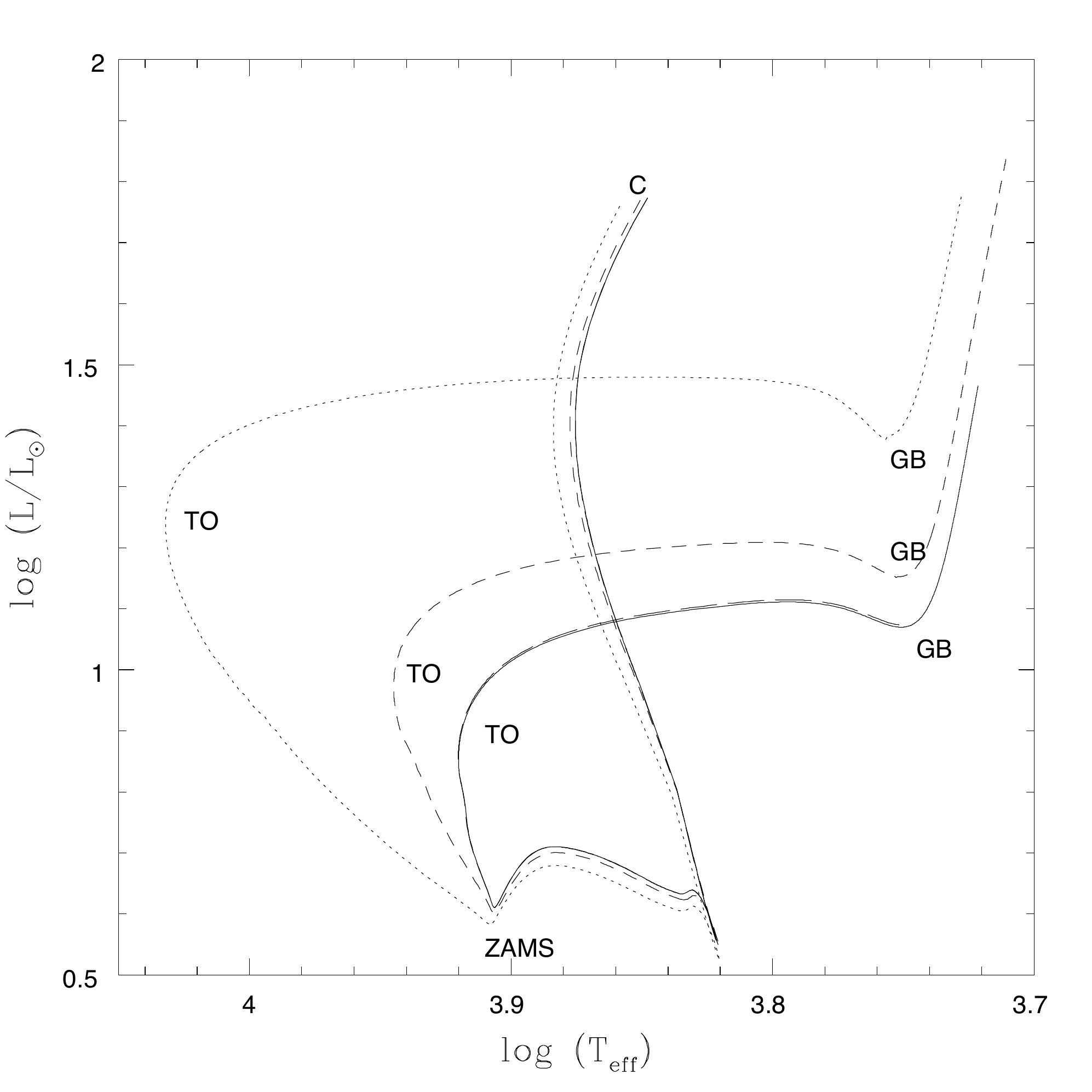}
\caption{Evolutionary tracks of the product of a collision between two 0.6 $\msun$ stars, after the angular velocity has been divided by a factor of 5 (dotted line), 10 (dashed line), 100 (long-dashed line), or 1000 (solid line). The end of collision phase is marked with `C',  and the zero-age main sequence (ZAMS), the turnoff\index{turnoff} (TO), and the base of the giant branch\index{red giant} (GB)  are also marked. (Figure 5 of \cite{Sills01}, reproduced by permission of the AAS)}
\label{fig:Rotating1}
\end{figure}

Rapidly rotating stars\index{rotating star} are subject to a number of instabilities\index{stellar instability} which can mix the stars on reasonably short timescales \cite{Maeder13}. These instabilities also modify their angular velocity profiles, often moving the star toward solid body rotation. At the same time, the collision products are contracting\index{contraction} towards the main sequence. If they do not have a way of losing their angular momentum, they must spin up, increasing the effectiveness of the mixing processes\index{mixing}. Because of the large initial angular momentum, and because the stars do not have any obvious way of losing their angular momentum (no discs, magnetic fields, or surface convective zones), the surface layers of the collision products are soon rotating more rapidly than the ``break-up" velocity\index{break-up velocity} at that radius, where the centrifugal force is larger than the local gravity. These layers become unbound and probably fly away from the star. However, they still carry very little of the total angular momentum. The star continues to contract and spin up, and loses more and more of its mass. Under these conditions, collision products could have a final mass much less than the turnoff mass in a cluster and therefore would not be identified as blue stragglers. However, during that short life, the rotational instabilities mixed the stars so efficiently that they could become very blue, as shown in Fig.~\ref{fig:Rotating1}. Here, the initial angular velocity of the same collision product has been arbitrarily divided by a factor of 5, 10, 100, or 1000, so that the collision products do not reach break-up velocities as they evolve to the main sequence and to show the effect of rotational mixing on the subsequent evolution. 

The evolution of rotating collision products, under assumptions of reasonable angular momentum loss prescriptions, was suggested by \cite{LeonardLivio95} and followed in detail by \cite{Sills05}. They assumed either a magnetic wind removed angular momentum, following the formalism applied to low mass main sequence stars\index{main sequence star} (e.g. \cite{Kawaler88}), or that the star was locked magnetically to a disc for a period of time, as assumed for pre-main sequence stars \cite{MattPudritz05}. Under either scenario, the collision products can lose an appropriate amount of angular momentum and therefore can live in the blue straggler region of the colour-magnitude diagram\index{colour-magnitude diagram} long enough to be observed there. However, they are still rotating fairly rapidly in this phase. Unfortunately, it is still too early to make a direct comparison between these rotation rates and the early observations of rotation rates of blue stragglers. Both the angular momentum loss prescriptions used by \cite{Sills05} assume that the collision product has a magnetic field\index{magnetic field} similar to that of the Sun and other normal, low mass (less than $\sim 1.2 \msun$) main sequence stars. We do not know anything about the magnetic properties of blue stragglers, and it is not clear that blue stragglers (with masses between 1 and $\sim1.8 \msun$ in globular clusters) should be modelled in the same way as lower-mass objects.

\begin{figure}
\sidecaption
\includegraphics[width=119mm]{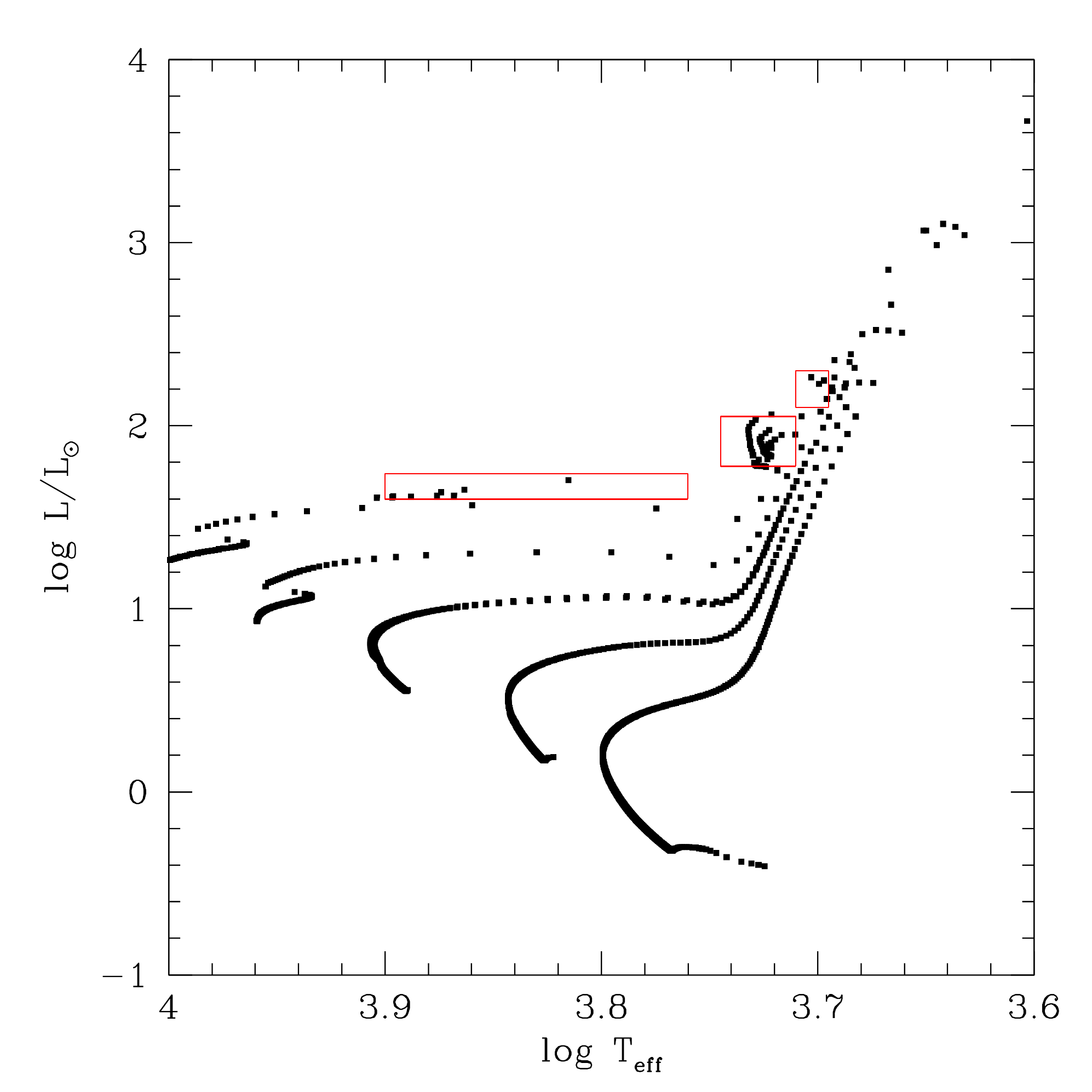}
\caption{Evolutionary tracks for all collision products for collisions which occurred 10 Gyr after the cluster was formed. The points are equally spaced at $10^7$ year intervals. The boxes outline the horizontal branch, E-BSS and asymptotic giant branch regions of the CMD. (Figure 11 from \cite{Sills09}, reproduced by permission of the AAS)}
\label{fig:EBSS}
\end{figure}

The models of collisions between main sequence stars are concerned only with blue stragglers, and therefore the evolutionary calculations are usually ended when the star leaves the main sequence. However, we might expect to see evolved blue stragglers in globular clusters, since the main sequence lifetime of a normal star with masses like blue stragglers ( $\sim1.2 \msun$) is only a few gigayears. The best place to look for these so-called E-BSS\index{evolved blue straggler} is on the horizontal branch\index{horizontal branch}, since massive stars should be brighter than the standard horizontal branch and quite red. The red giant\index{red giant} branch, on the other hand, is quite insensitive to the total mass of the star, and it is difficult to pick out unusual stars in this stage on the basis of photometry alone. Possible E-BSS have been identified in a few globular clusters. Sills, Karakas, \& Lattanzio\cite{Sills09} took collisional models and continued their evolution from the main sequence onto the asymptotic giant branch\index{asymptotic giant branch}. They found that the collision products are in the same place in the colour-magnitude diagram as the E-BSS during their horizontal branch phase, as shown in Fig.~ \ref{fig:EBSS}, and their horizontal branch lifetimes are consistent with the observed number of E-BSS, and almost independent of mass or initial composition profile. The ratio of evolved to main sequence blue stragglers in globular clusters also points to an average blue straggler main sequence lifetime of a few gigayears.

\begin{figure}
\sidecaption
\includegraphics[width=119mm]{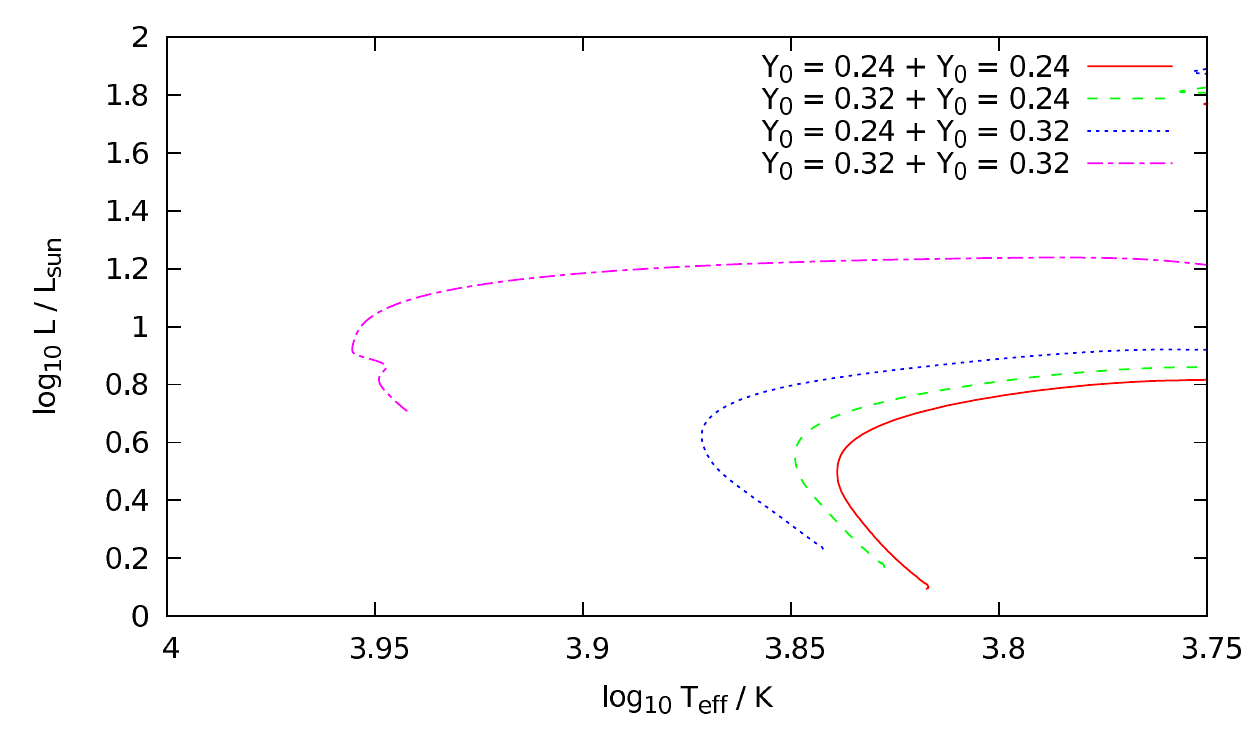}
\caption{Four evolutionary tracks of stellar collision products between a 0.6 $\msun$ star and a 0.4 $\msun$ star, with the two stars having helium abundances of either $Y=0.24$ or $Y=0.32$. The increased helium content modifies the position of the star in the HR diagram, and also reduces the main sequence lifetime for higher helium contents. (Figure 3 from \cite{GlebbeekSills10}, reproduced by permission of Oxford University Press on behalf of the RAS)}
\label{fig:MultiplePops}
\end{figure}

All the collisional models discussed above have assumed that the composition of the parent stars is the same, since the stars are expected to live in the same cluster. However, our recent understanding of globular clusters has changed. Increasingly, both photometric and spectroscopic evidence is pointing to at least two populations of stars in these clusters, one with a higher helium content than the other. Many stars have the helium content expected for a normal Population II\index{Population II star} composition ($Y \sim 0.23$), but a significant fraction of the stars could have helium as high as $Y ~ 0.4$ or so \cite{DAntona05}. These two populations are expected to both be formed early in the clusters' evolution. One implication of these two populations is that collisions could happen between high helium stars, or between one high and one low helium star. If a star has a higher than normal helium content\index{helium content}, it will be bluer and brighter than a normal star of the same mass, both of which are interesting properties for blue stragglers. Glebbeek, Sills \& Leigh \cite{GlebbeekSills10} began the investigation of how a parent population of varying helium content could affect the distribution of blue stragglers in the colour-magnitude diagram\index{colour-magnitude diagram}. They calculated evolutionary tracks for collision products between stars drawn from two populations (normal helium and helium-rich). A sample is shown in Fig.~\ref{fig:MultiplePops}. They found that for most clusters, the blue stragglers did not contain a large fraction of very high helium stars. However, NGC 2808\index{NGC 2808}'s blue stragglers were better fit by a model which included both populations. These findings are consistent with the expected amount of helium enhancement in the clusters studied, suggesting that a full understanding of blue straggler populations may also require us to pay close attention to the early evolution of globular clusters.

An additional property of blue stragglers that can be determined from stellar evolutionary models is the abundances\index{stellar abundance} of various elements at their surfaces. The predictions of collision models is that these stars should have surface compositions\index{surface composition} which are very similar to that of normal stars, since there is little to no mixing\index{mixing} during the collision\index{collision} process \cite{Lombardi96,Sills01}.  Lithium\index{lithium} should be depleted, because the very thin outer layers of low mass stars that are too cool for lithium burning will be ejected from the system during the collision. There may be some slight modification of carbon and nitrogen abundances of collision products on the red giant\index{red giant} branch \cite{Glebbeek08a} if the collision product contained some helium-rich material below the base of the convection zone, but blue stragglers still on the main sequence will not show this signature. 

When thinking about surface abundances, two effects need to be considered: first, what is the ``original'' surface abundance (i.e. immediately after the collision); and second, what happens to the abundances as the star evolves. There are a number of stellar processes which modify surface abundances, all of which are expected to occur in hot stars like blue stragglers under the right circumstances. Convection\index{convection} is a fast process which can fully mix the star to a particular depth, so if the chemical composition\index{chemical composition} of the star at the depth is different than the surface, the observed abundances will change. In rapidly rotating stars\index{rotating star}, as discussed above, various instabilities can mix quickly and to greater depths than convection. Thermohaline mixing\index{thermohaline mixing} occurs when material with a high mean molecular weight lies on top of a layer of smaller mean molecular weight. This process also mixes the two layers, but on a timescale which is longer than the convective timescale. If the stellar atmosphere is radiatively stable, then gravitational settling\index{gravitational settling} can cause the heavier atoms to sink relative to the lighter ones, resulting in a lower observed abundance of elements relative to hydrogen. Radiative levitation\index{radiative levitation} works in the opposite way, to push (typically) heavier elements up due to a tighter coupling between radiation and those elements. Both gravitational settling and radiative levitation are effective on similar timescales for interesting elements like carbon\index{carbon}, so any calculation which makes predictions of surface abundances must include both processes simultaneously.

\section{Binary Mass Transfer Models}
\label{sec:BinaryModels}

The other likely mechanism for creating blue stragglers is through mass transfer in a binary system. This mechanism is distinct from stellar collisions in that the two stars are in a stable orbit, and the mass transfer\index{mass transfer} occurs because of some internal evolution of the stars or the system. Normally, one star's radius increases due to normal stellar evolution and the star fills its Roche lobe\index{Roche lobe overflow}, or the binary orbit shrinks (e.g., due to angular momentum loss via gravitational radiation or a magnetized wind) so that mass transfer will occur. In a cluster, the binary orbit could also be modified by interactions. However, the models of the mass transfer event and the subsequent evolution of the blue straggler are always done in isolation without considering the environment surrounding the binary. 

Unlike collisions, mass transfer can be a very slow process. Under some circumstances, mass can be transferred from one star to another on timescales that are comparable to the nuclear timescale\index{nuclear timescale} of the individual stars. To model such systems, stellar evolution codes\index{stellar evolution code} which can simultaneously evolve both stars are used. The main code used in the literature for this type of work is a modified version of Eggleton's stellar evolution code, described in detail in \cite{Nelson01}. Rather than assuming that the mass of each star is fixed in time, an additional outer boundary condition is introduced so that mass can be lost if the star overfills its Roche lobe. An equivalent boundary condition is used when a star loses mass through a stellar wind, and in some instances, both a wind and Roche lobe overflow are modelled \cite{ChenHan08}. A fraction $\beta$ of the material lost by the primary is accreted\index{accretion} onto the secondary, where $\beta=1.0$ for fully conservative mass transfer. The orbit of the binary is recalculated using the new masses and the assumption that the material carries the specific angular momentum\index{angular momentum} of mass-losing star. 

These codes are very robust when the mass transfer rate is reasonably low, but if the donor star is unstable to dynamical mass transfer, they can break down. For a full discussion of the conditions for stable and unstable mass transfer, see Chap.8 in this volume. The binary evolution\index{binary evolution} codes typically treat dynamical mass transfer by setting a maximum upper limit on the total amount of mass that can be transferred in any given timestep (or a maximum mass loss rate\index{mass loss rate}). If the parameters of the star and binary system are such that the mass transfer rates reach this maximum, one of two things are done. Either the mass loss rate is arbitrarily set to a high, but still stable, mass transfer rate (e.g. \cite{ChenHan08b}), or the two stars are assumed to merge and a structure for the merger product is assumed (e.g. \cite{ChenHan08}). Unfortunately, various groups do not agree on the assumed structure of a dynamical merger product. Chen \& Han \cite{ChenHan08} argue that a dynamical merger product will have the same structure as a stellar collision product, while  Lu, Deng \& Zhang \cite{Lu10} assume that mergers are fully mixed.

As described in Chap. 8, Roche lobe overflow\index{Roche lobe overflow} mass transfer is usually divided into three cases, following \cite{Kipp67} and \cite{lau70}. Case A\index{Cases A, B, C of mass transfer} mass transfer occurs when the donor is on the main sequence and is typically stable on nuclear timescales. Case B mass transfer occurs a little later, after the main sequence but before core helium ignition, and Case C is mostly when the donor is on the asymptotic giant branch\index{asymptotic giant branch}. Case B and Case C mass transfer can be unstable on either dynamical\index{dynamical timescale} or thermal timescales\index{thermal timescale}, depending on the structure of the envelope of the star, primarily the depth of the convective zone. All three cases are expected to contribute to the formation of blue stragglers, although the relative importance of each channel is not yet well understood.

\begin{figure}
\sidecaption
\includegraphics[angle=90,width=119mm]{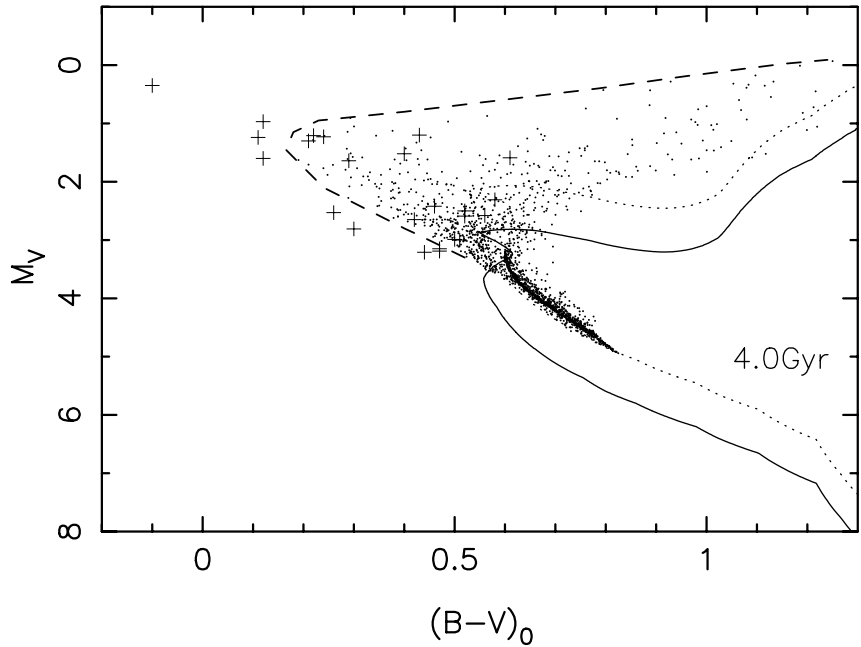}
\caption{A colour-magnitude diagram\index{colour-magnitude diagram} for a 4 Gyr old system showing the locations of binaries which are currently undergoing mass transfer as small dots. The solid line is a 4 Gyr isochrone, and the dotted line is that isochrone moved brighter by 0.75 magnitudes, to show where equal-mass binaries would lie. The dashed line gives the upper edge of the binary population. The crosses are the observed positions of the blue stragglers in M67. (Figure 5 from \cite{Tian06}, reproduced with permission \copyright\,ESO)}
\label{fig:CaseACurrent}
\end{figure}

Models of binary mass transfer products are done self-consistently if the stars do not fully merge. For example, Tian et al. \cite{Tian06} follows the evolution of a 1.4 + 0.9 $\msun$ binary which transfers mass while the 1.4 $\msun$ star is on the main sequence, continues as that star becomes a giant, but ends when the mass ratio reverses, leaving a blue straggler and a helium white dwarf\index{helium white dwarf} binary system. They also calculate a full grid of such case A mass transfer systems, with various initial masses and orbital separations. At an age of 4 Gyr (to compare to the open cluster M67\index{M67}), they find that many blue stragglers are still undergoing mass transfer. Both the primary and the secondary can still contribute significantly to the total luminosity of the system, and so these objects cover a very large area in the colour-magnitude diagram\index{colour-magnitude diagram} (Fig.~ \ref{fig:CaseACurrent}). They are all above a line which is 0.75 magnitudes brighter than the Zero Age Main Sequence, and populate the area between this line and the giant branch due to blends with the second star in the system. 
\begin{figure}
\sidecaption
\includegraphics[angle=90,width=119mm]{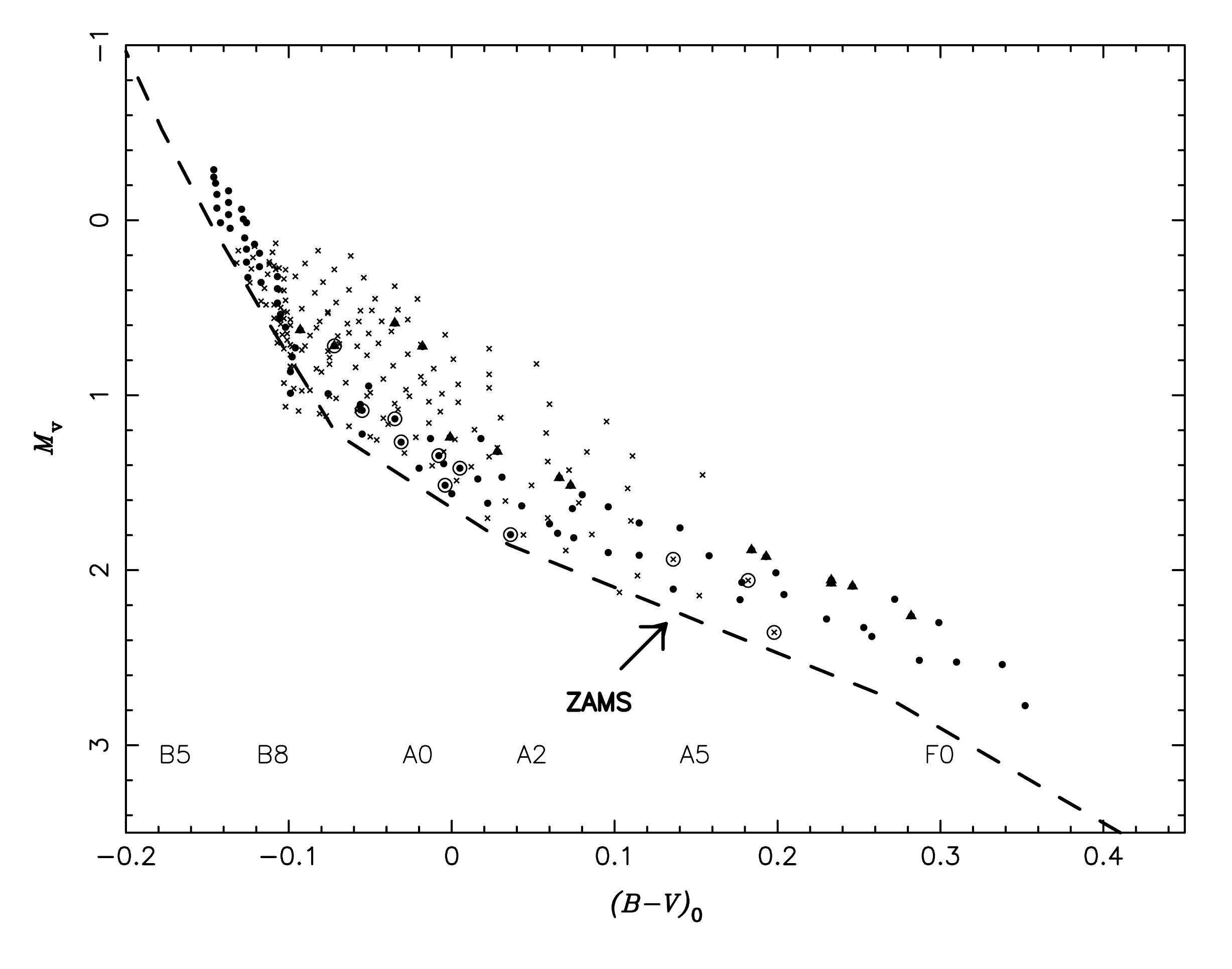}
\caption{Location of case A binary mergers in the colour-magnitude diagram\index{colour-magnitude diagram}. Dots and crosses represent systems which evolved quickly and slowly to contact\index{contact binary}, respectively. The triangles are models with very low hydrogen content. The open symbols are observed blue stragglers from a number of open clusters. (Figure 1 from \cite{ChenHan08}, reproduced by permission of Oxford University Press on behalf of the RAS)}
\label{fig:CaseAMergers}
\end{figure}

To study mergers, Chen \& Han \cite{ChenHan08} looked at case A mass transfer\index{mass transfer}\index{Cases A, B, C of mass transfer} in systems that result in a merger (based on the initial periods and masses of the binary system) after a rapid (thermal timescale) or slow (nuclear timescale) evolution to contact.  The evolution of these binary systems before contact was followed using the Eggleton code, under the assumption that the material from the primary is distributed homogeneously over the surface of the secondary. The evolution was stopped when contact was reached, and they assumed that the merger product has the core of the primary at the centre, and then a fully-mixed envelope which consists of the material from the secondary mixed with the envelope of the primary. These objects are then evolved using a single-star evolution model. Because the merger products can have a wide range surface abundances but are still single stars in various stages of main sequence evolution, they all lie within about 1 magnitude of the Zero Age Main Sequence (Fig.~ \ref{fig:CaseAMergers}). However, the hydrodynamic process of merging is not followed in detail, and these calculations assume that the merger product is in thermal equilibrium\index{thermal equilibrium} before the evolution begins. If the thermal contraction phase of a merger product is long, or if the object is like a collision product and does not fully mix the secondary into the envelope, then these evolutionary tracks need to be modified. Currently, there are not many hydrodynamic simulations of the merger of contact binaries, so we cannot definitively predict the appropriate structure of a post-mass-transfer merger. 

\begin{figure}
\sidecaption
\includegraphics[angle=90,width=12cm]{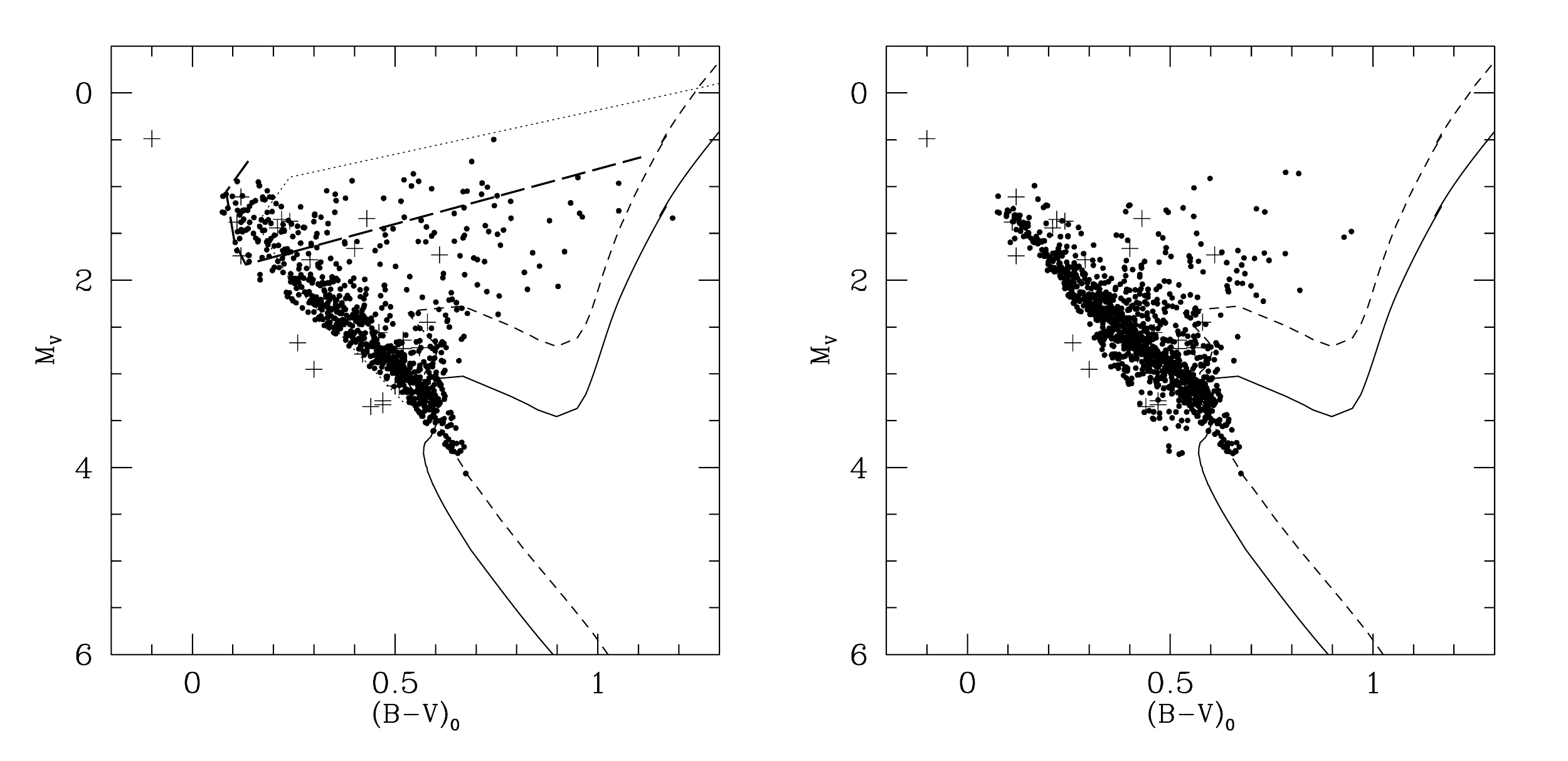}
\caption{A colour-magnitude diagram of a population of 4 Gyr old binaries undergoing case B mass transfer (solid dots) and blue stragglers observed in M67 (crosses). The conservative mass transfer parameter $\beta$ is set to 1.0 in the left panel and 0.5 in the right. The solid line is a 4 Gyr isochrone\index{isochrone}, and the small dashed line is the same isochrone shifted by 0.75 magnitudes, representing the locus of equal-mass binaries. The dotted line gives the upper edge of the case A binary population, taken from \cite{Tian06} (Fig.~\ref{fig:CaseACurrent}). The long-dashed line gives the lower bound of case B binaries with primary masses between 1.4 and 1.5 $\msun$.  (Figure 6 from \cite{Lu10}, reproduced by permission of Oxford University Press on behalf of the RAS)}
\label{fig:CaseB}
\end{figure}

Models of stable case B mass transfer\index{Cases A, B, C of mass transfer}\index{mass transfer} can only be calculated when mass transfer begins when the primary is in the Hertzsprung gap\index{Hertzsprung gap} (e.g., \cite{Lu10}). As the primary moves up the giant\index{red giant} branch, it develops a larger convective envelope and quickly becomes unstable to mass transfer\index{mass transfer} on a dynamical timescale\index{dynamical timescale}. These systems will result in a merger or a common envelope\index{common envelope} system, whose evolution is not well understood \cite{Ivanova13}. The stable case B mass transfer systems tend to be slightly bluer and brighter than the case A systems (compare Fig.~\ref{fig:CaseB} with Fig.~\ref{fig:CaseACurrent}) but still cover much of the area between the line 0.75 magnitudes above the ZAMS, and the giant branch. Few systems can lie below that line, unless the conservative mass transfer parameter $\beta$ is less than one. 

There are very few models of case C mass transfer\index{Cases A, B, C of mass transfer} \index{mass transfer} applied specifically to blue stragglers. Chen \& Han \cite{ChenHan07} looked at the critical mass ratio\index{mass ratio} for stable mass transfer in binaries appropriate for M67\index{M67}, and conclude that case C could provide a viable formation mechanism for the long-period blue stragglers in that cluster. Since the surface chemistry of AGB stars\index{AGB star} is significantly different than that of their less-evolved red giant cousins, we might expect that blue stragglers formed through case C mass transfer would show evidence of anomalous chemistry (s-process elements, for example). AGB stars also have less mass to donate to their companion to form a blue straggler, since their core is more massive than that of a red giant. And case C mass transfer must occur in wide binaries, because the system did not undergo mass transfer on the red giant branch. Other than that, the expectation is that the overall evolution will be similar to objects which are undergoing case B mass transfer. 

Very little is known about the rotational properties of the products of mass transfer\index{mass transfer}. The models discussed above do not include stellar rotation\index{stellar rotation}. It is reasonable to assume that merger\index{merger} products will be rotating rapidly immediately after the merger, since the angular momentum\index{angular momentum} of the orbit will have to be absorbed by the star. The subsequent evolution will depend on the structure of the object immediately after the merger, and any star which remains rapidly rotating for a substantial period of time will become more and more mixed\index{mixing}. Since we do not see blue stragglers that have anomalously high rotation rates or extremely blue locations in the CMD, we can assume that there is some efficient angular momentum loss process for binary mergers, analogous to what must occur in stellar collision products. 

Surface abundances, particularly of CNO elements\index{CNO elements}, are thought to be a tracer of a binary mass transfer formation mechanism \cite{ChenHan04}. If the mass transfer process exposes material which has been processed through CNO burning\index{CNO burning} (from the core or a hydrogen-burning shell of a red giant), then we should see evidence for mass transfer on the surface of the blue straggler. Just like in stellar collisions, however, it is important to also consider the subsequent evolution of the surface abundances due to internal mixing. Chen \& Han \cite{ChenHan04} looked at thermohaline mixing\index{thermohaline mixing} in merger products, and concluded that the CNO signature of mass transfer would remain but that the position of the star in the colour-magnitude diagram\index{colour-magnitude diagram} would be affected slightly. Glebbeek et al. \cite{Glebbeek10} compare a particular collision and a binary merger from the same two parent stars, and conclude that the CNO abundances of the two stars will be similar, but that lithium\index{lithium} may provide a better discriminant. A larger investigation of parameter space is required before we can feel confident that abundances along can pinpoint a formation mechanism for a blue straggler.

\section{Parametrised Models}
\label{sec:ParameterizedModels}

In the context of stellar dynamics, it is very time-consuming to calculate detailed stellar evolutionary\index{stellar evolution code} models for all stars in a system (a cluster or a galaxy). In the case of galaxies\index{galaxy}, the system is non-collisional and so stars can be treated as point masses without compromising the scientific questions that are being asked. For collisional systems such as dense star clusters\index{star cluster}, however, the physical radii of the stars as a function of time becomes important in determining the dynamical state of the system at later times. In other words, collisions\index{collision} can happen, and if we are to model the system correctly, direct physical collisions must be included. In addition, it can be important to track the mass of each star over time, since mass loss from stars can significantly modify the cluster potential \cite{Hills80}. We also want to know the luminosity and temperature of each star to create synthetic colour-magnitude diagrams\index{colour-magnitude diagram}. In order to capture all the stellar quantities that a stellar evolution code would give, but without doing the detailed evolution calculations, a number of codes have been developed which provide parametrised stellar evolution.  In other words, the radius, temperature, luminosity, and mass of a star of a given initial mass can be determined at any time $t$ by making use of functional fits to evolutionary tracks. The next obvious step for such codes was to combine these stellar evolution recipes with the binary evolution recipes that have been developed for binary population synthesis codes. Processes such as Roche Lobe overflow\index{Roche lobe overflow} mass transfer, common envelope evolution\index{common-envelope evolution}, wind mass transfer\index{wind mass transfer}, and orbital energy loss by gravitational wave radiation are all parametrised. Criteria for each process to occur, and importantly the outcome of each process (in terms of the new masses, evolutionary status, and orbital elements of the binary) are given as simple formulae. These codes can also include a prescription for the outcome of direct stellar collisions. 

The three main parameterized binary and stellar evolution codes used in the recent literature are {\tt SeBa}\index{SeBa code} \cite{SPZ96}, {\tt BSE}\index{BSE code} \cite{Hurley02}, and {\tt StarTrack}\index{StarTrack code} \cite{Belczynski08}. {\tt BSE} is the most commonly used code, and it has been implemented into a number of stellar dynamics codes such as {\tt NBODY4/NBODY6/NBODY7}\index{NBODY code} \cite{Aarseth12}, {\tt Cluster Monte Carlo}\index{Cluster Monte Carlo code} \cite{Chatterjee10} and {\tt MOCCA}\index{MOCCA code} \cite{Hypki13} and is also available as a module in {\tt AMUSE}\index{AMUSE code} \cite{SPZ13}. {\tt SeBa} is an integral component of the {\tt STARLAB}\index{STARLAB code} code, and {\tt StarTrack} is primarily a binary population synthesis code but has been combined with simple dynamic simulations of clusters to model the dynamical modification of binary populations in clusters \cite{Ivanova05}. 

The parametrised approach is very good for determining the overall properties of a stellar system. It does a better job than any dynamical simulation which does not include binary or collisional processes, and allows for direct comparison to observations. However, because the prescriptions for many of the binary evolutionary processes and collisions are simplified, we cannot use these parameterized codes to compare directly to individual blue stragglers. Even the details of the blue straggler populations are not treated correctly. For example, observations of blue straggler stars in NGC 188\index{NGC 188} (see Chap. 3) shows that {\tt BSE} over-predicts the number of common envelope\index{common envelope} systems. Comparison of {\tt BSE} results with detailed stellar evolution models of stellar collisions \cite{Glebbeek08a} and with the observed positions of blue stragglers in the colour-magnitude diagrams\index{colour-magnitude diagram} of globular clusters\index{globular cluster} \cite{Sills13} show that the {\tt BSE} treatment of collisions\index{collision} overestimates their lifetime. {\tt BSE}, which was originally formulated in the early 1990s, still assumes that collision products are fully mixed, despite the numerous papers showing that mixing is not expected either during or after the collision. Therefore, we caution against using only parametrised codes when trying to predict the properties of individual blue stragglers or blue straggler populations in clusters.
	
\section{Future Directions}
\label{sec:7}

Models of individual blue stragglers have become more sophisticated over the past two decades. Both colliding stars and binary mass transfer have been modelled in detail using hydrodynamical techniques. Binary mergers\index{merger}, also known as coalescence, are usually assigned an assumed post-merger structure. All these objects are then evolved using a stellar evolution code, which, in the case of the mass transfer models, consistently treats the material which is leaving one star and accreting on the other. The resulting evolutionary tracks are compared to the observed positions of blue stragglers in colour-magnitude diagrams to try to constrain their formation mechanism(s). 

There are still a number of unanswered questions and additional observations that are yet to be investigated in the context of detailed evolutionary models. The modelling of collision products is the more mature field of the two main formation mechanisms. Even though binary mass transfer is probably a more common process, and is relevant for many objects beyond blue stragglers, there is more work to be done here. The parameter space of possible binaries is much larger and has not been fully explored. We probably understand fairly well how a mass-gaining star evolves, but the evolution of stars which fully merge after an episode of mass transfer is less clear. Are they indeed fully mixed, or does the core of the primary retain its integrity? More detailed models, although difficult, are necessary to fully understand this important contribution to blue stragglers. 

One place where collisional models could be improved is through a more careful treatment of appropriate parent stars. While the idea of multiple populations, with multiple helium contents may be specific to globular clusters, it may be important to include different chemical compositions in those situations. It is also important to remember that blue stragglers can have a fairly long lifetime, so they can have formed long ago when the parent stars were less evolved, and potentially more massive. The same can be said for mass transfer products, since stable mass transfer can proceed for many gigayears in some cases and so the amount of mass and its composition may not be what one would expect if we only consider stars near the current turnoff mass. 

Many blue stragglers are known pulsators\index{pulsating star}. They are known as SX Phoenicis stars\index{SX Phe star}, and are the low metallicity, low mass counterparts of $\delta$ Scuti stars\index{$\delta$ Scuti star}. They are radial pulsators, and lie in the same instability strip\index{instability strip} as RR Lyrae stars\index{RR Lyrae star} and Cepheids\index{Cepheid}. They are found in globular clusters\index{globular cluster} and dwarf galaxies\index{dwarf galaxy}, and most lie on the expected period-luminosity relation \cite{Cohen12}. There is a population of sub-luminous SX Phe stars in globular clusters which may have higher helium\index{helium} abundances. Whether these abundances are related to the internal structure of individual blue stragglers, or to the possibility of multiple populations in globular clusters, remains to be seen. Only a few groups have invested in detailed models of SX Phe stars \cite{Santolamazza01, Templeton02}, and both were more concerned with determining the location of the instability strip. We know from studies of the more common radial pulsators (RR Lyraes and Cepheids) that pulsation modes can be a powerful tool for determining fundamental stellar parameters such as their mass and chemical composition. A detailed comparison between observed pulsation modes and models may provide a way of distinguishing between various formation mechanisms, as long as those mechanisms produce stars which are substantially different in some fundamental property.

There are two pieces of physics that stellar modellers typically try to avoid: rotation\index{stellar rotation} and magnetic fields\index{magnetic field}. Unfortunately, it seems that both effects are going to be important for both stellar collisions and binary mergers. Angular momentum\index{angular momentum} loss, probably mediated by a magnetic wind or magnetic locking to a disc is needed. In particular, magnetohydynamic simulations of collisions and of binary mergers may be needed to determine what happens to any primordial magnetic field of the parent stars. Rotational mixing\index{mixing} is also likely to be important to the subsequent evolution of blue stragglers if they rotate even slightly faster than normal for most of their main sequence lifetime. The current rotation rates of blue stragglers are still puzzles that have not been addressed in much detail. Modellers will have to bite the bullet and include at least one of these difficult physical effects, and probably both.

% Use the \index{} command to code your index words
%

%
%\begin{acknowledgement}
%Write something nice here about the workshop? Thank ESO for their hospitality? Thank Evert for his slides. 
%\end{acknowledgement}

%

\backmatter%%%%%%%%%%%%%%%%%%%%%%%%%%%%%%%%%%%%%%%%%%%%%%%%%%%%%%%
%\appendix
%\include{appendix}
%\include{glossary}
\printindex

%%%%%%%%%%%%%%%%%%%%%%%%%%%%%%%%%%%%%%%%%%%%%%%%%%%%%%%%%%%%%%%%%%%%%%

\end{document}